\documentclass{aastex}          %% The manuscript based on AASTeX v5.x
\usepackage{spr-astr-addons}
\begin{document}

   \title{Evolution and instabilities of disks harboring super massive black holes}

    \author{Anna Curir \altaffilmark{1}, Valentina
de Romeri \altaffilmark{1}, Giuseppe Murante \altaffilmark{1}}

%	\offprints{A. Curir}
     \altaffiltext{1}{INAF-Osservatorio Astronomico di Torino. Strada Osservatorio 20
 -10025 Pino Torinese (Torino). Italy.   e-mail:curir@oato.inaf.it}
     \date{}
  
\begin{abstract}
The bar formation is still an open problem in modern astrophysics.  In this
paper  we present numerical simulation performed with the aim of analyzing the growth
of the bar instability inside stellar-gaseous disks, where the star formation
is triggered, and a central black hole is present.
The aim of this paper is to point out the impact of such a central massive
black hole on the growth of the bar. 
We use N-body-SPH simulations of the same isolated disk-to-halo mass systems
harboring black holes with different initial masses and different energy
feedback on the surrounding gas. We compare the results of these simulations
with the one of the same disk without black hole in its center. We make the
same comparison (disk with and without black hole) for a stellar disk in a
fully cosmological scenario. 
 A stellar bar, lasting 10 Gyrs, is present in all our simulations.\\
The central black hole mass has in general  a mild effect on the ellipticity of the bar
but it is never able to destroy it.  The black holes grow in different way
according their initial mass and their feedback efficiency, the final values of the
velocity dispersions and of the black hole masses are near to the
phenomenological constraints.  
\end{abstract}

%\authorrunning{}
%\titlerunning{Black holes in barred disks}

\maketitle

\section{Introduction}

In a series of recent papers \citet{Cu06}, \citet{Cu07}, \citet{Cu08} we have
shown that the bar formation inside disk galaxies is triggered also by the
cosmological scenario and not only by the classical instability of a self
gravitating disk. We obtained such result by analyzing the growth of a bar
instability in exponential disks, embedded in a Dark Matter (DM) halo which
evolves in a cosmological context. We used purely stellar disks as well as
disks containing both gas and stars. In the latter case, we studied the effect
of only allowing the gas to radiatively cool, and we then examined the impact
of the star formation. Our embedding technique allowed us to vary the baryon
to DM mass ratio onto the {\it same} DM halo. In almost all the cases we
studied, including the classically stable ones, we noticed the formation of a long--lasting bar.  The onset of the bar instability, in the latter cases, appears to be driven
by the triaxiality and by the dynamical evolution of the DM halo. For a few
cases, cooling gas proved to be able to stabilize the gas, but such effect
disappeared when star formation was turned on, because gas converts in stars
before a stabilizing, dense, central knot can form.Therefore the problem of
inhibiting the bar formation is still an open astrophysical problem, since one
third at least of disk galaxies are not barred.

The observational data indicate that massive central black holes exist in disk
galaxies as well as ellipticals. Moreover black holes masses are correlated
with host galaxies properties; \citet{Mag98} has concluded that
the median black holes mass is 0.006 of the bulge mass and \citet{Kor01} correlated the black hole mass to the bulge velocity
dispersion.  Such a large mass concentration inside the galaxy could affect
its structure.  It is known indeed that a central mass concentrations (CMC) can
destroy a bar (see e. g. \citet{Bou05}, \citet{Cu07} ). The effect of the
presence of a massive black hole (BH) at the center
of a disk galaxy could therefore mimic the stabilizing action of a knot of
dense, cold gas which forms if gas is allowed to radiatively cool but not to
form stars.

The influence of central BHs on the dynamical evolution of bars in
disk galaxies has been examined mainly by \citet{she04} and \citet{Hoz05} in
a pure non dissipative scenario and using as model galaxies isolated
systems. They results give very high values ($10^{8.5}$) the minimum mass necessary for bar
dissolution, higher than the inferred one for local spirals.

In this paper we want to treat the same problem using simulations of stellar-
gaseous disk with star formation.  We will evolve the disk model as isolated
system and in a cosmological scenario.  We will discuss the evolution of the
bar ellipticity and semi-major axis of the same disk endowed with different
initial BH  mass. The BH  is accreting during the
evolution. Its accretion is regulated by a suitable feedback
parameter(\citet{DiMat03}), which states a coupling between a percentage of
the accretion radiating energy and the gas heating.  We will also show the
role of this feedback parameter and its interplay with the initial BH
mass, the formation of the CMC and with the star formation.

Finally, we will investigate the relation between the decrease of the bar
ellipticity,the accretion time and the feedback.  Moreover we will compare our
simulations with some phenomenological issues as the Kormendy black hole
mass-bulge velocity dispersion relation.

The plan of the paper is the following. In Section 2, we summarize our recipe
for the initial $disk+halo$ system and present the star formation recipe.  In
Section 3, we present our simulations, and in Section 4 we point out our
results.  The parameters related to the bars formed in the new and the old
stars and to the global, old+new, stellar populations are given in this
section.  In section 5 there are  phenomenological implications and section 6 is devoted to our discussion and conclusions.

\section{ Method}

In the cosmological simulations, we embed a gaseous and stellar disk inside a
cosmological halo selected in a suitable way, as to be able to host a disk
galaxy, and follow its evolution inside a cosmological framework: a
$\Lambda$CDM model with $\Omega_{m}=$0.3, $\Omega_{\Lambda}=$0.7,
$\sigma_8=$0.9, $h=$0.7, where $\Omega_{m}$ is the total matter of the
Universe, $\Omega_{\Lambda}$ the cosmological constant, $\sigma_8$ the
normalization of the power spectrum, and $h$ the value of the Hubble constant
in units of 100$h^{-1}$ km\,s$^{-1}$Mpc$^{-1}$. The halo is extracted from the
cosmological simulation at a given redshift (we chose $z=2$). The
stellar+gaseous disk is embedded in the halo, in equilibrium with its
gravitational potential as in the isolated case; the system is then
re-inserted in the full cosmological simulation and evolved.  A detailed
description of our method of producing the cosmological scenario where the
disk is evolved was given in \cite{Cu06}.

As far as the isolated simulations is concerned, we embed a gaseous and
stellar disk inside a DM halo, having a radial density profile of the NFW form
\citet{Nav96} and follow its evolution for 10 Gyrs.  The
halo has a mass of $ 1.36\times 10^{11} h^{-1} M_{\odot} $  and initial concentration
c=4.5. The disk is set in equilibrium with the gravitational potential of the
DM halo, but not {\it vice-versa}; as a consequence, there is a transient
period in which the halo responds to the disk immersion. In Curir (2006) we
showed that our embedding procedure is not triggering, {\it per-se}, a bar
instability, and as well, no instability is triggered because of our chosen
numerical resolution.

The units of our simulations are:
$ 10^{10} h^{-1} M_{\odot} $ for the mass, $1 h^{-1}   kpc$ for the length,
$0.98   h^{-1} Gyrs$ for the
time.

\subsection{The baryonic disk}

The spatial distribution of particles follows the exponential surface density
law: $\rho =\rho_0\exp -(r/r_0)$ where r$_0$ is the disk scale length,
$r_0=4h^{-1}$\, Kpc, and $\rho_{0}$ the surface central density.  The disk is
truncated at five scale lengths with a radius: R$_{disk}=$20$h^{-1}$\,Kpc.
Its mass is $1.9 \times 10^{10} h^{-1} M_{\odot}$.\\  This value of mass has been
  chosen and 
 employed in the previous papers as the critical one for the bar instability:
the disk/halo mass ratio for this case is in the instability range from the
classical point of view, the presence of a cold gas knot originated by
gas cooling does stabilize it against the bar formation, and it
becomes unstable again if the gas is allowed to form stars.

To obtain each disk particle's
position according to the assumed density distribution, we used the rejection
method \cite{Pre86}.  We used 56000 star particles and 56000 gaseous particles
to describe our disk.  The masses of gas, stellar and DM particles are
  respectively $7 \times 10^{4} h^{-1} M _{\odot }, 2.8 \times 10^{5} h^{-1}  M _{\odot },
  1.194 \times 10^{6} h^{-1} M_{\odot } $. The (Plummer equivalent) softening length, the same for
DM, gas, and star particles  and black hole, is $ 0.5 h^{-1}$\,kpc. 

In the cosmological case we embed the disk in the high-resolution cosmological
simulation, at redshift 2, in a plane perpendicular to the angular momentum
vector of the halo and in gravitational equilibrium with the potential.  Its
center of mass corresponds to the position of the DM particle having the
minimum value of gravitational potential.  The initial redshift corresponds to
10.24 Gyrs down to $z=0$ in our chosen cosmology. 

During the evolution, new star particles are formed from the gaseous
particles.  We refer to such new component as `new star component', whereas we
will call `old star component' the non dissipative particles present in the
disks at $z=2$ (at $t=0$ in the isolated cases).

In the disk center of mass resides a BH which is represented as a
collisionless, heavy mass point. It is able to swallow the gas in its
neighborhood.  The feedback parameter$\epsilon_f$ assigns a certain percentage of the
accretion energy to be thermally coupled with the gaseous component.  This
feedback regulates the BH accretion, which, without such effect, is
catastrophic.  The default value of such parameter is $ {\epsilon_f} =0.05$ (see
below), but we will explore in the paper other different values. We used the
BH/energy feedback model of \citet{DiMat03}.
This model hypothesizes that
the gas accretion is limited by the Eddington rate: 
$$
 \dot{M} = \frac{4 \pi G M_{BH} m_p}{\epsilon_r \sigma_T c} 
$$ 

where $M_{BH}$ is the BH mass,
$m_p$ the proton mass, $\sigma_T$ the Thompson cross section and $\epsilon_r$ the
radiative efficiency. We follow \citet{DiMat03} and set $\epsilon_r=0.1$. A
fraction $\epsilon_f$ of the energy emitted by the BH couples with the surrounding
gas, giving an energy output rate of $E_{feed} = \epsilon_f \epsilon_r M_{BH}
c^2$. The parameter $\epsilon_f$ is the feedback parameter.

\subsection{Star formation recipe}

We use the sub-grid star formation prescription by Springel \& Hernquist
(2003). In it, when a gas particle overcomes a given density threshold its gas
content, is considered to reside in a multi-phase state, that corresponds to
the equilibrium solution of an analytical model describing the physics of the
multi-phase interstellar medium. Such a solution gives an effective
temperature for the gas particle, obtained as a weighted average of the
(fixed) temperature of the cold phase and of the hot phase.  The temperature
of the hot phase is set by supernova feedback and by the efficiency of cloud
evaporation. The resulting inter stellar medium (ISM) has an effective
equation of state $P_{eff}(\rho)$, which is stiffer than isothermal and
prevents Toomre instabilities even when a large amount of gas is present.
When this prescription is used, the star formation rate agrees with the
Schmidt law in disk galaxies, as obtained e.g. by Kennicut (1998).  The
effective temperature drives hydrodynamical interactions of the gas
particles. Moreover, the model consistently gives a star formation rate, which
is used to spawn a star particle from the gas one on the basis of a stochastic
prescription; the initial mass function adopted is the Salpeter's one. The
star-forming from the gas has the same position and velocity as the gaseous
particle.

\section{Simulations}

We performed 11 simulations of a disk+halo isolated system, and 2 Cosmological
simulations, where the disk was embedded in our selected cosmological halo.
We exploited the non-public version of the parallel Tree+SPH N-body code
GADGET-2 \citep{Spri05} (courtesy of V. Springel).
%(except one case, simulation 5 of table 1, which had only% The cell-opening criterion is based on the absolute truncation
%error of the multiple expansion for particle-cells interaction (see Springel {\it et al.},
%2001, for details). 
The simulations run on the  CLX computers located at the CINECA computing
center (BO, Italy ).  The configurations have been evolved for  10.44 Gyrs $(t_{fin}) $.

\begin{table*}
\caption{ Simulations:   o.s.=old stars , n.s.= new stars. Masses are in units of
  $10^{10} h^{-1} M_{\odot}$, lengths in $h^{-1}$ kpc, times in Gyrs}
\label{cosmsimtable}
\centering
\begin{tabular}{c c c c c c c c c c }
\hline\hline
%\noalign{\smallskip}
%\noalign{\smallskip}
N  & BH $m_{ini}$ &  $\epsilon_f$ & $\epsilon$ (o.s.)&  $\epsilon$ (n.s.)  & BH $m_{fin}$ & a$_{max}$ (o.s.)
& $a_{max}$ (n.s.)&   $CMC-BH_{frac}$ & $t_{BH=10^8}$ \\  
\hline
 i1 &  $ 10^{-6}$ & 0 & 0.747 & 0.75 & 0.18 & 6.37 & 1.6 & 0.23 - 0.72& 6.289 \\
 12 &  0.001 &  0 & 0.43 & 0.44 & 0.23 & 6.28 & 6.27&  0.3 - 0.76 & 2.05\\
 i3 & 0.005  & $10^{-4}$ & 0.845 & 0.85 & 0.021 & 7.7 & 1.2  & 0.19 - 0.11 & 0.993  \\
 i4 &  $10^{-5}$ & 0.05 & 0.8 & 0.81 & $6.8 \times  10^{-5}$  & 7.94 & 1.5& 0.22 -
 $10^{-4}$ & $t>t_{fin}$ \\
 i5 &  $10^{-4}$ & 0.05  & 0.69 & 0.71 & $1.15 \times 10^{-4}$  & 6.16 & 6.6 &
 0.14 - $8.2 \times 10^{-4}$ & $t>t_{fin}$\\
 i6 & $ 10^{-4}$ & 0.01 & 0.74 & 0.82 &  $2. \times 10^{-4}$ & 3.45 & 1.31&
 0.15 - $1.3 \times 10^{-3}$ & {$t>t_{fin}$}\\
 i7 &  0.001 & $10^{-5}$&  0.58 & 0.49 &  0.21 & 6.59 & 3.73 & 0.16 - 1.3 & 0.1667 \\
 i8 & 0.001 &  $5. \times 10^{-5}$ & 0.77 & 0.69 & 0.027 & 8.29 & 2.49& 0.19 - 1.42 & 1.389   \\
 i9 & 0.001 &  $10^{-4}$ & 0.73 & 0.75 & 0.014 & 8.16 & 1.96& 0.18 - 0.078 & 5.49 \\
 i10 & 0.001 &  0.001 & 0.83 & 0.83 & 0.0024 & 6.72 & 1.21& 0.16 - 0.015 & $t>t_{fin}${}\\
 i11 & 0.001 &  0.01 & 0.74 & 0.85 & 0.0018 & 7.8 & 1.36 & 0.17 - 0.01& $t>t_{fin}${} \\
 c1 & 0.001  & $5. \times 10^{-5}$ &  0.53 & 0.477&   0.0065    &  8.18&    4.08&   0.15 - 0.04 & $t>t_{fin}${}  \\
 c2 &   0   &    0 &    0.476   & 0.48 &   0   &  7.2&    3.8&  0.1 & $t>t_{fin}${} \\
\hline
\end{tabular}
\end{table*}

The main parameters and the final properties of our set of simulations are
listed in Table \ref{cosmsimtable}. We kept fixed the disk mass and the gas to
stellar mass ratio, and varied the initial BH mass and the feedback strength
$\epsilon_f$. We focused on the effect of the presence of a BH on the bar
instability, and refer the reader to our previous works \citet{Cu06},
\citet{Cu07}, \citet{Cu08} for a study of the impact of other significant
parameters of our disk+halo system on such instability.

As a geometrical global measure of the bar strength, we defined the value of
the ellipticity, $\epsilon=1-b/a$, (where $a$, $b$ are the semi-axes of
the contour) chosen on the iso-density level where such a
value is maximum.

We defined the {\it bulge} as the total mass contained inside a radius
centered on the new stars center of mass and corresponding to a
minimum thickness of $4 h^{-1} kpc$.\\  The bulges in our simulations are
  ``pseudobulges''  as defined in  \citet{Kor}. The initial disk is bulgeless
  and the bulge is formed by secular instability and accreted by the formation
  of the new stars. In our simulations the timescale of the  formation of these pseudobulges is
  3-4 Gyrs. The final masses consisting of old and new stars   are between $6 \times 10^9  h^{-1} M_{\odot} $ and $7 \times 19^9  h^{-1} M_{\odot}$. 

In Table \ref{cosmsimtable} we present the simulation number (I column), the
initial mass of the BH in code units (II col.), the feedback value,
i.e. the percentage of accretion energy coupled with the gas heating (III
col.).  Moreover, as final values we present the maximum ellipticity for the
old star system(IV col.) and for the new star (V col.), the final mass of the
accreted BH (VI col.), the major axis corresponding to
the maximum ellipticity for the old stars (VII col.)and for the new stars
(VIII col.), the  central mass  concentration CMC defined as the mass of all
the components (stars, new stars and gas) inside a radius of $1.5 h^{-1}$  kpc followed by
the mass fraction of BH to the CMC (IX col.) and the time needed for the accreting
BH to reach the value $10^8  h^{-1} M_{\odot}$ .

\section{Results}

Our main result is that in all of our simulations, a stellar bar is
still living at the final time in the old star component.  The new
star component at the end of our runs is arranged in a
bulge component which presents a more or less barred shape depending on
the initial mass of the BH and on the feedback. Overall, the
presence of a BH is never able to quench the stellar bar, not
even when its mass is significantly larger than allowed by observations.

\begin{figure*}
\includegraphics[width= 8cm]{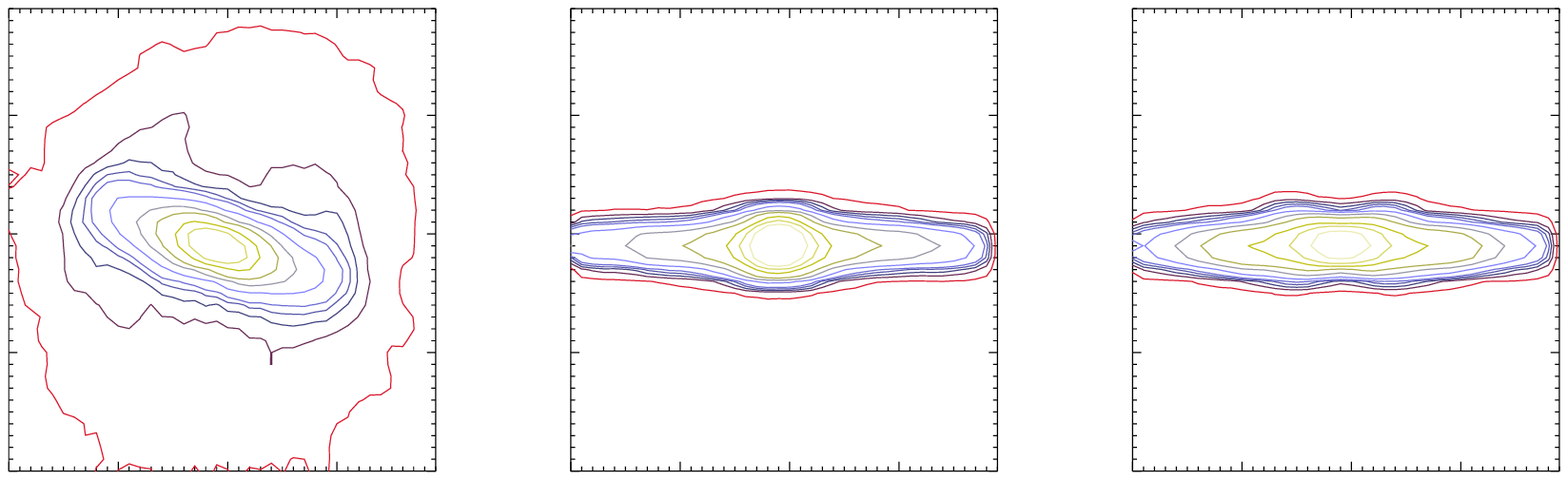}
\includegraphics[width= 8cm]{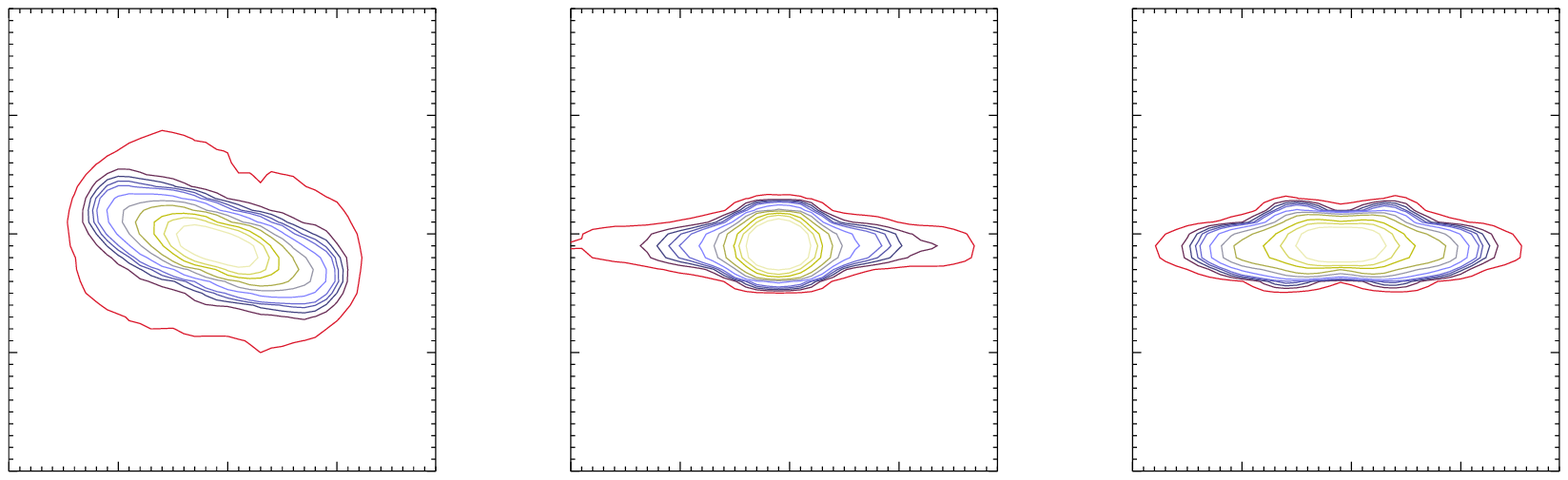}
\caption{iso-density contours  of old stars in the planes xy, xz and yz (left three panels) and new stars (right panels) for simulation i8,  at the end of the evolution}
\label{fin}
\end{figure*}

An example of the final configuration is shown in Fig. \ref{fin} for simulation
i8. Fig \ref{fin} has been constructed  with the same
box-size, number of levels, and density contrast  as in Papers 1, 2 and 3.\\

\begin{figure*}
\includegraphics[width= 8cm]{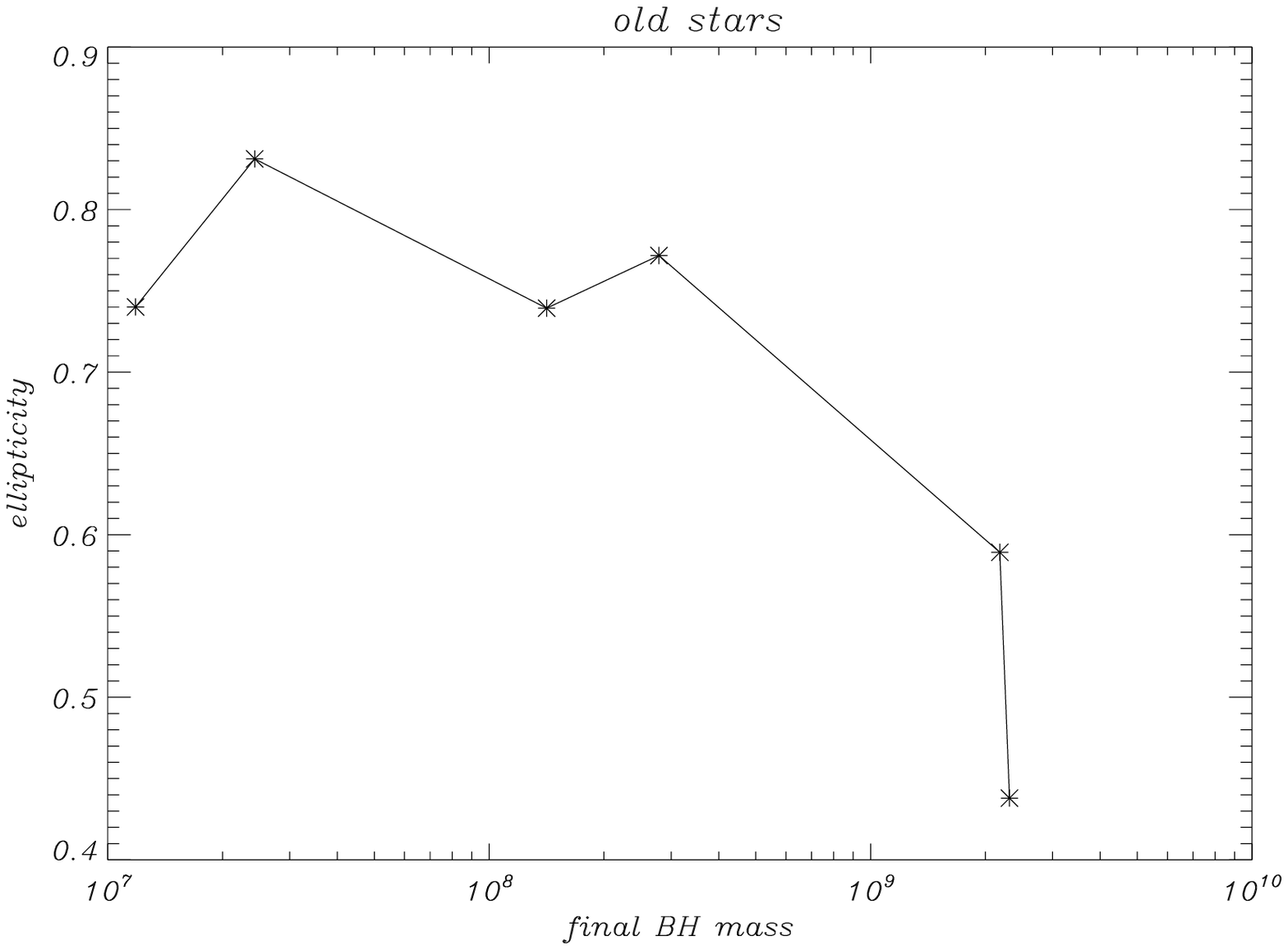}
\includegraphics[width= 8cm]{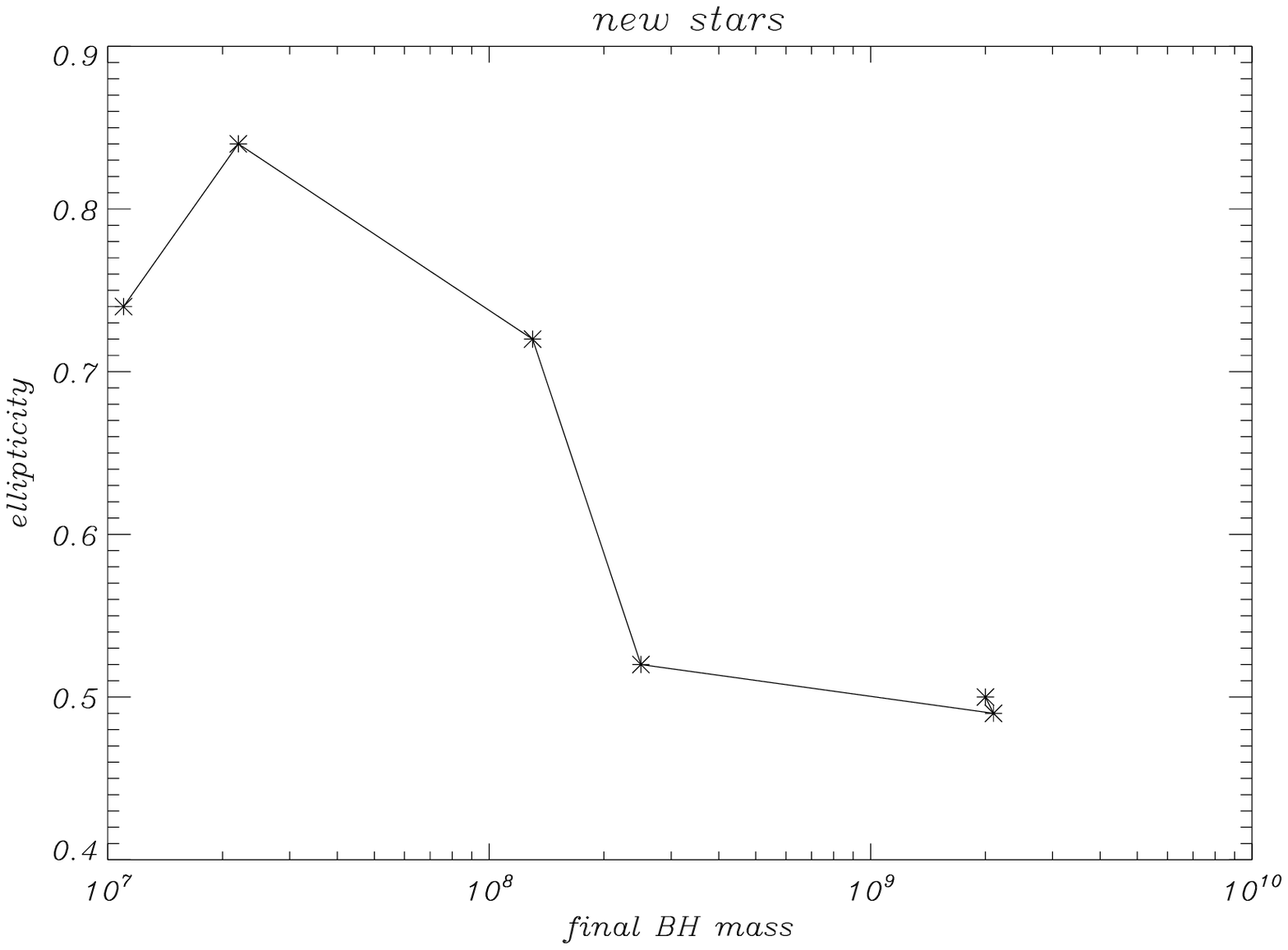}
\caption{Ellipticity  as function of the final mass (in solar masses) of the
 BH for the simulations with the same BH's initial mass ($10^7  h^{-1} M_{\odot}$).}
\end{figure*}

In Fig. 2 we show the value of the ellipticity of the bar at the final
time, as function of the final mass of the BH.  Only results
for simulations having the same initial masses of the BH
($10^{7} h^{-1}  M_{\odot}$) are shown here.  The final BH mass 
is a direct function of the feedback parameter, the larger one
corresponding to a simulation with no BH feedback.  This
diagram show a clear, if obvious, correlation  for the masses larger than
$10^8 h^{-1} M_{\odot}$: more massive black
hole-weaker bar.  In particular, the point of lower ellipticity in
such figure correspond to the final time of simulation i2, where the
feedback parameter is zero and the BH mass is $2.3
\times {10^{9}}h^{-1}  M_{\odot} $, a value higher than the minimum mass
requested by \citet{Hoz05} for a complete bar destruction.  In our
case the bar instability is not totally inhibited, but the ellipticity
is the weakest one.

 In Fig. 2 the peak in the ellipticity    corresponding to a feedback $\epsilon_f = 0.001$
(simulation  i10)  is created by the competition between two effects:  the star
formation which is working in favor of the bar (and which is weakened by the
BH feedback)   and the increase of the
central BH mass + CMC, which is inhibiting the bar.  The star formation rate 
increases moving from the first to the second point of diagram 2 (because the
BH feedback  decreases), it is
stationary going to the third point  (but the central BH mass + CMC  has increased
enough to weaken the ellipticity) and it decreases again after.

\begin{figure}
%\epsscale{0.5}
%\figurenum{1a}
%\begin{center}
\includegraphics[width=7cm]{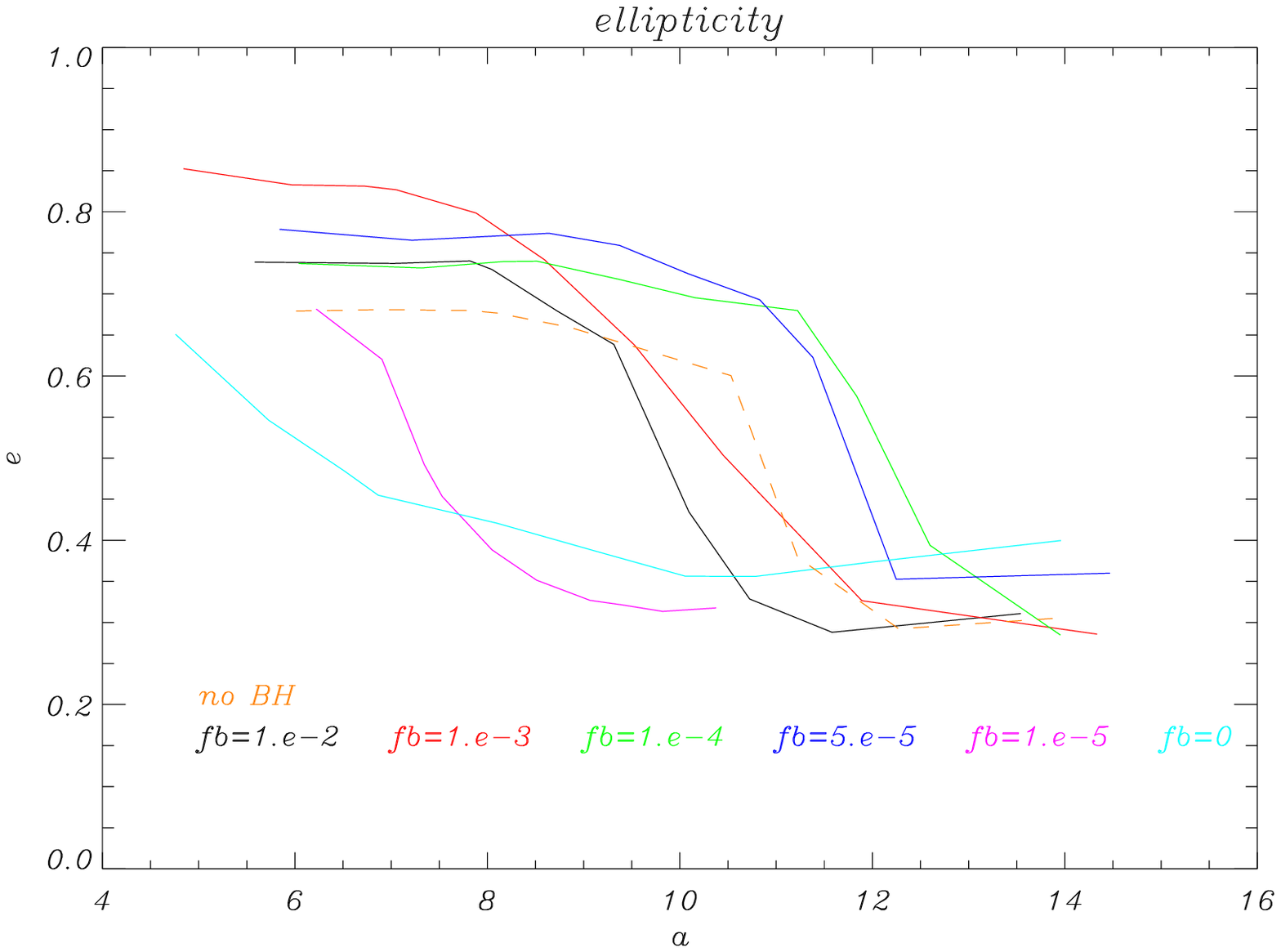}
\includegraphics[width=7cm]{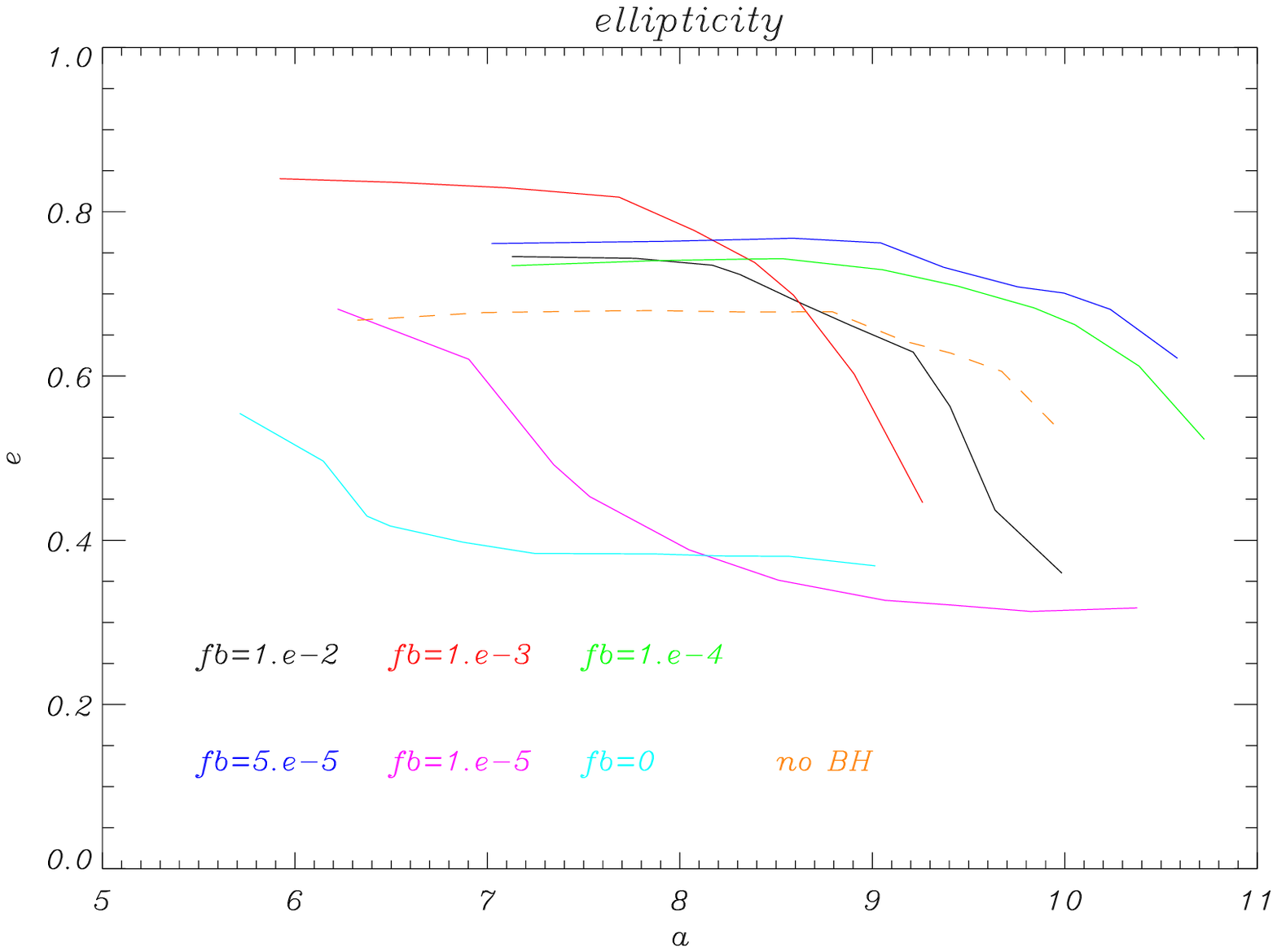}
%%\includegraphics[width=10cm]{f6bis.eps}
%\end{center}
\caption{ Profiles of the ellipticity at the final time for the  set of
 simulations  with the same initial BH mass $10^7 h^{-1} M_{\odot}$ and
 different feedback.  The profiles are for the old stars component (upper
 panel) and for the new one (lower panel)} 
% $Q_b$ (dotted line) and ellipticity (full line) of our
%more massive disks (i.e. disk--to--halo mass ratio
%0.33); $Q_b$ (dot--dashed line) and ellipticity (dashed line) of our
%less massive, DM-dominated, disks (i.e. disk--to--halo mass ratio
%0.1).
%}
%\label{strength}
\end{figure}

\begin{figure}
%\epsscale{0.5}
%\figurenum{1a}
%\begin{center}
\includegraphics[width=7cm]{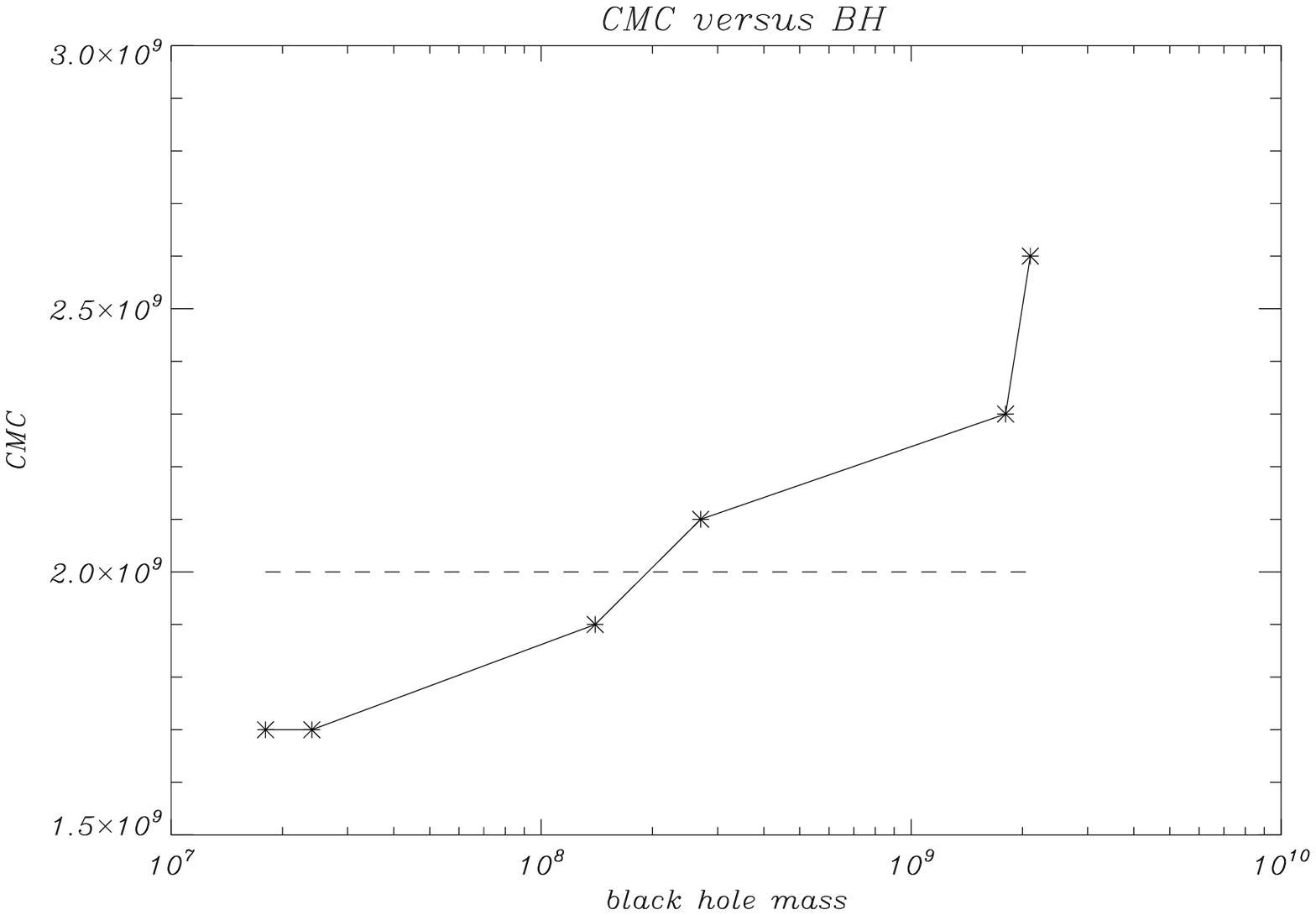}
%%\includegraphics[width=10cm]{f6bis.eps}
%\end{center}
\caption{ Final central mass concentration (in solar masses) versus final BH mass (in solar masses), for the cases having the
  same initial BH mass.  The horizontal line indicates the value of the final
  CMC in the case of the disk without BH.} 
% $Q_b$ (dotted line) and ellipticity (full line) of our
%more massive disks (i.e. disk--to--halo mass ratio
%0.33); $Q_b$ (dot--dashed line) and ellipticity (dashed line) of our
%less massive, DM-dominated, disks (i.e. disk--to--halo mass ratio
%0.1).
%}
%\label{strength}
\end{figure}

 More details about the impact of the BH on the ellipticity are in Fig. 3, where we  see the final ellipticity profile, as a function
  of the semi-major axes of the iso-density contours for the
  simulations of Fig. 2.

 In Table 1, we  see that the values of the CMC are different for
  different simulations and in most of the cases they are bigger, or much
  bigger than the BH final mass.  Therefore one can deduce that the effect of
  the black hole in the disk center is not really a direct gravitational impact on
  the bar, but more an effect of modulating and, for the more massive BH and
  lower feedback values, enhancing the CMC formation,
  which, in turn, has an effect on the bar ellipticity. In Fig. 4 we have
  shown the direct correlation between the final BH mass and the CMC: the more
  the black hole is massive, the higher is the CMC. In Fig.4  we present
  only the cases starting with the same initial BH mass: as in Fig. 2, 
  more massive BHs correspond to lower values of the feedback. In the same figure,   we
  show an horizontal line: the value of the CMC for the case of the disk without BH. It is
  evident that for the higher values of the feedback (first three points of the diagram), the value of the CMC
is reduced by the gas heating, when compared with the CMC of the same disk without a
central  BH.  The difference in CMC between cases with strong feedback and
weak one is almost completely due o the new stars component and to the gas content.
  
We also compared the effect of having different evolutionary patterns
of our central BH on the onset of the bar, i.e. different initial BH
masses with different feedback efficiencies.  With this aim, we looked
for the time $t$ at which the mass of the BH is equal to the
value $10^{8} h^{-1} M_{\odot}$, independently of the initial BH
mass. We then plotted in Fig 5 the ellipticity as a function of the
time $t$ needed for the BH to  reach such a mass value.

We can disentangle the effect of the bar evolution by comparing the
bar ellipticities measured in the same disk, at the same times, in a
simulation without a central BH (dotted line in Fig.5).

This comparison shows that a too fast or too slow
growth of the BH results in a small effect on the bar ellipticity. A significant change of ellipticity, with respect to the
simulation without the BH, only happens if the BH takes from 2
to 5 Gyrs to reach a mass of $10^8$ M$_\odot$.  Of course, the
subsequent evolution of the bar {\it does} depend on the mass of the
BH; a BH which quickly reaches our reference mass and
then continues to grow, will have a larger effect on the bar
ellipticity at the final time, as shown in Figure 2.

%This comparison shows that for old stars there is a threshold lapse
%of time ( $\approx$ 3 gyrs) which is critical for the accreting black
%hole to drop the value of the ellipticity of the bar.  if the time
%accretion to get the mass 0.01 reaches 6 Gyrs the bar instability
%prevails on the black hole effect, and the bar acquires almost the
%same strength and length as in the case without the black hole.\\ For
%the new stars on the other hand we notice that the bar ellipticity is
%lower than the ellipticity of the new stars bar if the black hole is
%not present in the accretion range between 3 and 6 Gyears. \\

\begin{figure*}
\includegraphics[width= 8cm]{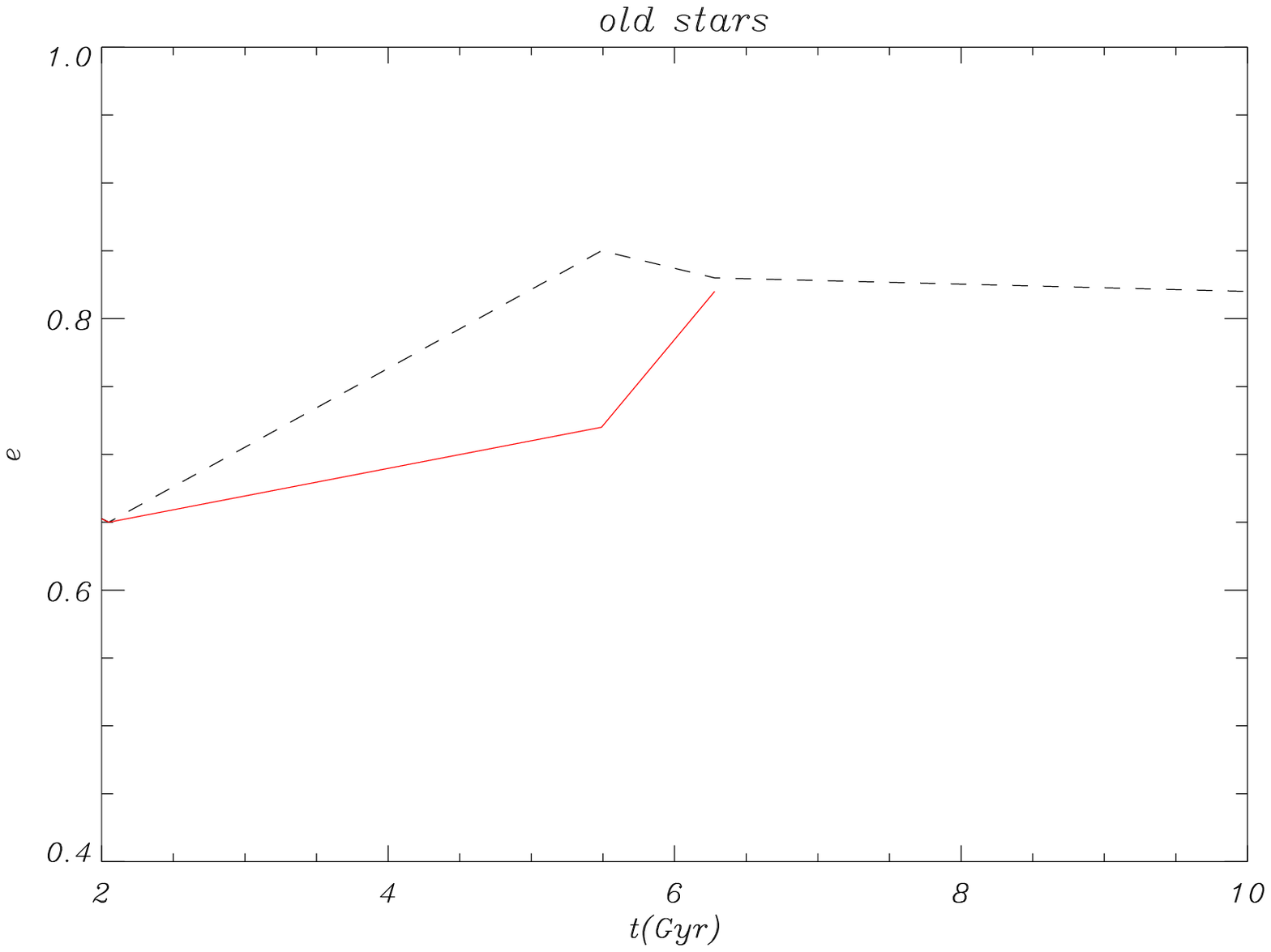}
\includegraphics[width= 8cm]{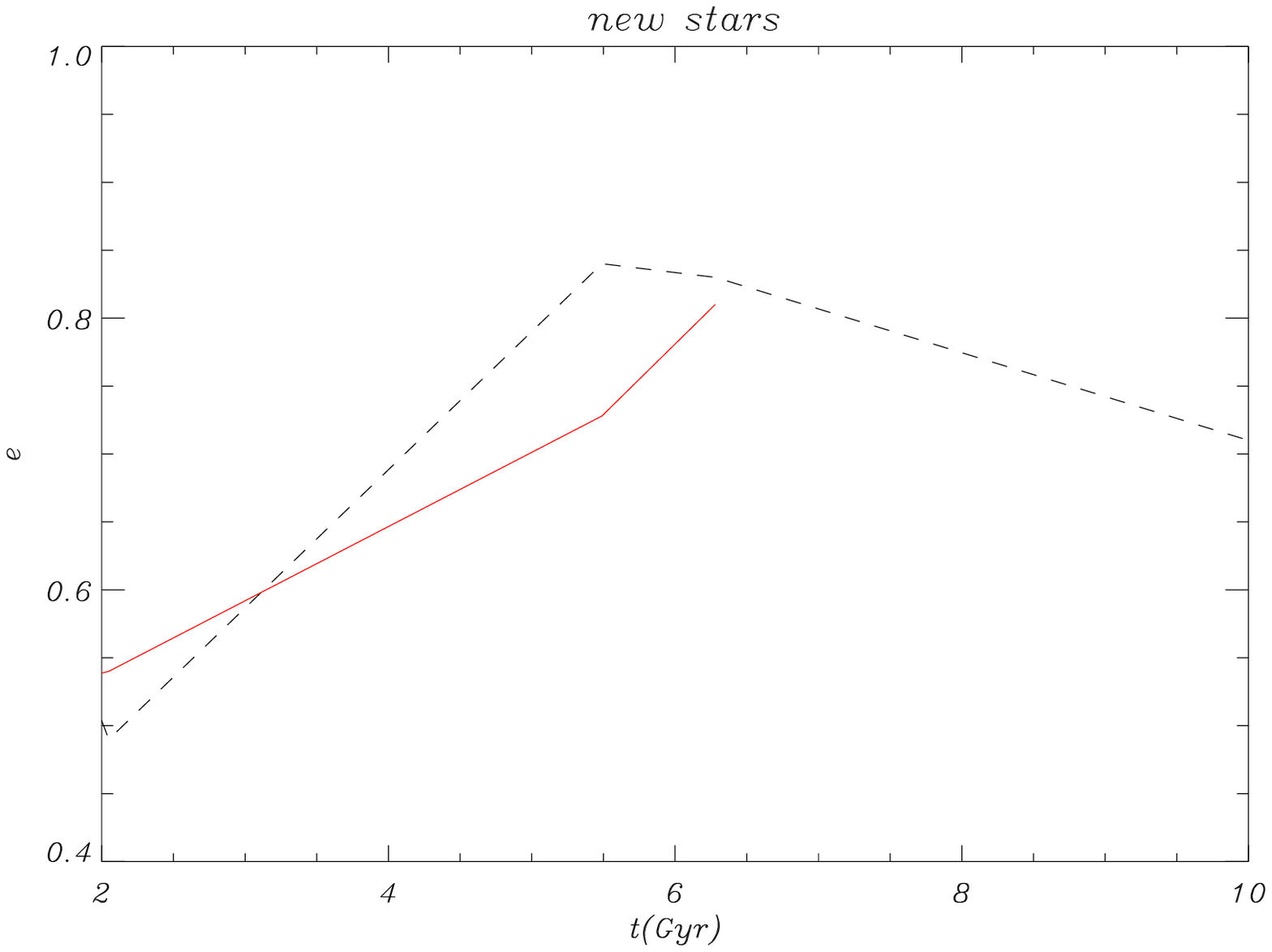}
\caption{Ellipticity of the bar when the accreting BH reaches the value
  $10^{8} h^{-1} M_{\odot} $ , as function of the time accretion (solid line). Ellipticity at the same
  time for a disk without central BH (dashed line).  The ellipticity is
evaluated for old star (left  panel) and new stars (right panel)}
%(right panel).}
%\label{gas_z1_03}
\end{figure*}
%\begin{figure}
%\epsscale{0.5}
%\figurenum{1a}
%\begin{center}
%\includegraphics[width=7cm]{8285f13.ps}
%%\includegraphics[width=10cm]{f6bis.eps}
%\end{center}
%\caption{ Behaviour of the bar strength and ellipticity at z$=0$ for our set of
%cosmological simulations  with increasing gas fraction. 
% $Q_b$ (dotted line) and ellipticity (full line) of our
%more massive disks (i.e. disk--to--halo mass ratio
%0.33); $Q_b$ (dot--dashed line) and ellipticity (dashed line) of our
%less massive, DM-dominated, disks (i.e. disk--to--halo mass ratio
%0.1).
%}
%\label{strength_time}
%\end{figure*}
%\begin{figure*}
%%\includegraphics[width= 6cm]{8285f12a.ps}
%\includegraphics[width= 16cm]{ell_cfr_tempo_simsenzaBH_ns-1.ps}
%\caption{same as in \ref{strength_time} but for new stars}
%\label{strength_time_ns}
%\end{figure*}

%We provide in Fig 5 and 6 the behavior of the ellipticity profiles of the old
%stars (Fig. 5 ) and new stars at the accretion times considered in
%Fig. 4. One can see that for both the disk+BH and the plain disk cases the
%ellipticity increases with the evolution in the internal region of the bar.
%Such an increase is slightly milder in very central radii if the black hole is
%present, especially at t=5.59 for the old stars.
%For the new stars, the ellipticity appears lower at earlier times and at all
%radii  if the black hole is present.\\

In Fig.6 and 7 we provide the profiles of the {\it differences} of the
ellipticities with and without the BH, for
the old (Fig. 6) and the new (Fig.7) stars at the same times as in
Fig. 5., i.e. when the BH mass reaches the value $10^8$ M$_\odot$.
These differences are always negatives, for the new stars except for
t=2.05 (case i2 in Tab. 1), meaning that for all the other times, the bar is stronger at
all radii if the BH is not present. 

For the old stars profiles the trend is less evident: at some times we
have radii in the disk where the ellipticities of the bar containing
the BH is slightly higher. But it is evident that the differences
are always negative at all times as far as the internal radii is
concerned.  We interpret this as the confirm that the effect of the
BH is to decrease the strength of the inner bars.  This is
consistent with the findings of \citet{Holley}
which however only considered the stellar component in a triaxial system
 and used an
adiabatically growing potential to simulate the presence and growth of
the central BH.  They noticed that the major impact of the BH on the triaxiality of the
galaxy involves at maximum the $10\% $ inner mass of the galaxy.
We  estimated indeed that the major impact on the variation of the  ellipticity
is on the  {\it inner}
iso-density contour, containing $\approx 10 \%$ of the mass, when the BH is present
with respect to our control case. 

These results are qualitatively confirmed even if we evaluate the ellipticity
using the inertial tensor inside our  stellar system.

\begin{figure}
\includegraphics[width= 8cm]{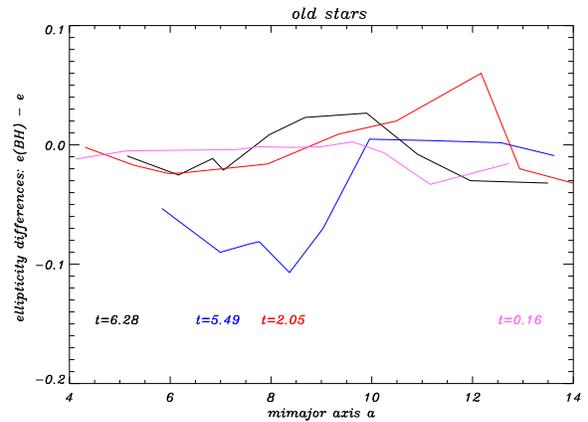}
\caption{Differences between ellipticity profiles of the stellar bar (old stars) with (e(BH)) and
  without (e) the BH, when the accreting BH reach the value
  $10^8 h^{-1} M_{\odot}$, the different colors correspond to different accretion times.}
%(right panel).}
%\label{gas_z1_03}
%\end{figure}
%\begin{figure}
%\epsscale{0.5}
%\figurenum{1a}
%\begin{center}
%\includegraphics[width=7cm]{8285f13.ps}
%%\includegraphics[width=10cm]{f6bis.eps}
%\end{center}
%\caption{ Behaviour of the bar strength and ellipticity at z$=0$ for our set of
%cosmological simulations  with increasing gas fraction. 
% $Q_b$ (dotted line) and ellipticity (full line) of our
%more massive disks (i.e. disk--to--halo mass ratio
%0.33); $Q_b$ (dot--dashed line) and ellipticity (dashed line) of our
%less massive, DM-dominated, disks (i.e. disk--to--halo mass ratio
%0.1).
%}
\label{ell_prof}
\end{figure}

\begin{figure}
\includegraphics[width= 8cm]{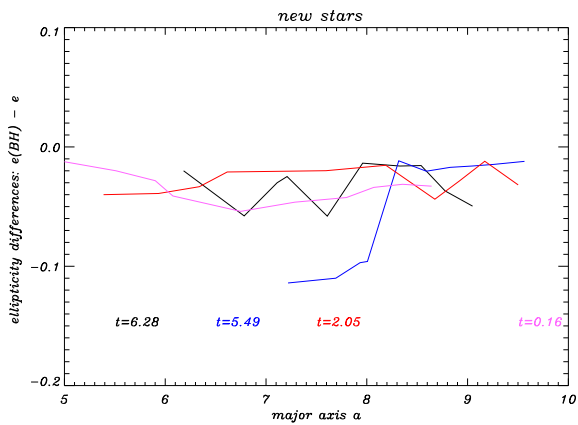}
\caption{Same as in the previous figure, but  for the new stars.}
%(right panel).}
%\label{gas_z1_03}
%\end{figure}
%\begin{figure}
%\epsscale{0.5}
%\figurenum{1a}
%\begin{center}
%\includegraphics[width=7cm]{8285f13.ps}
%%\includegraphics[width=10cm]{f6bis.eps}
%\end{center}
%\caption{ Behaviour of the bar strength and ellipticity at z$=0$ for our set of
%cosmological simulations  with increasing gas fraction. 
% $Q_b$ (dotted line) and ellipticity (full line) of our
%more massive disks (i.e. disk--to--halo mass ratio
%0.33); $Q_b$ (dot--dashed line) and ellipticity (dashed line) of our
%less massive, DM-dominated, disks (i.e. disk--to--halo mass ratio
%0.1).
%}
\label{ell_prof2}
\end{figure}
\subsection{Cosmological cases}
\subsubsection{bar strength}

We performed two cosmological simulations, one with a BH having a mass
$M_{BH}= 10^7 h^{-1} M_\odot$ (c1) and another without the BH (c2).  We chose 
$ {\epsilon}_f = 5. 10^{-5}$ as our feedback parameter. We used the embedding procedure
described in \citet{Cu06}.  The values of the
ellipticities show a very mild impact of the BH on the onset
of the bar instability and its evolution.  In Table 1 we reported the
final ($z=0$) value of the bar ellipticity, which is very similar in
the two cosmological cases both in the old and in the new stellar
component. Note that, for the cosmological run, we used a quite high
BH initial mass and a low feedback parameter. Such parameter choice is
the one which shows to be the most effective in dumping the bar
instability in our isolated cases (case i8 in Table 1). Even so,
the presence of the central BH has practically no inhibiting effect on
the bar instability.

\subsubsection{ DM concentration}

In this section we investigate if the BH has an impact on the
concentration parameter of the haloes.  We then compare three different
haloes: the halo from our original DM-only simulation, selected for the disk
immersion; the same halo containing the disk, and the same halo containing the
disk and the BH.

In the table \ref{concentration} we show the behavior of the NFW
concentration parameter as a function of the redshift $z$ in the three
cases.  It is known that the baryonic matter has an influence on the
halo concentration increasing it by an amount that \citet{Lin06}, e.g,
estimated to be around $10\%$. Moreover, the concentration evolves
approximately linearly with redshift (see e.g. \citet{Bull01}); at
redshift $z=2$ we expect a concentration parameter of the order of one
third that found at $z=0$.  Our DM halo shows a concentration
evolution which is evolving more than linearly: it increases by a
factor $\approx 4$ from $z=2$ to $z=0$.  In the Table, we also show
the increase in the concentration of the DM halo due to the presence
of the baryonic disk only (case c1).   This increase is higher than the 
 expected  $10\%$.  On the other hand, from the table it is clear that the
presence of a BH has practically no effect on the concentration
parameter.

\begin{table*}
%\centering
\caption{Halo concentrations}
\label{concentration}
\begin{tabular}{c c c c }
\hline\hline
%\noalign{\smallskip}
%\noalign{\smallskip}
z  & halo & c1 & c2\\  
\hline
 2 & 4.5 & 4.5 & 4.5 \\
 1 & 8.8 & 12.8 & 12.9 \\
 0 & 19.7 & 29.1 &	29.9  \\
\hline
\end{tabular}
\end{table*}

%\begin{figure}
%\begin{center}
%%\includegraphics[width= 6cm]{8285f12a.ps}
%\includegraphics[width= 8cm]{fig7_rew.ps}
%\end{center}
%\caption{Comparison of the halo concentrations at different times of evolution:
%the full line describe the pure DM cosmological halo; the dashed line, the
%halo concentration when the disk is embedded; the dotted line the halo
%concentration with the disk and the BH are present}
%\label{concent}
%\end{figure}

\section{Kormendy relation}
 
\begin{figure}
%\begin{center}
\includegraphics[width= 6cm]{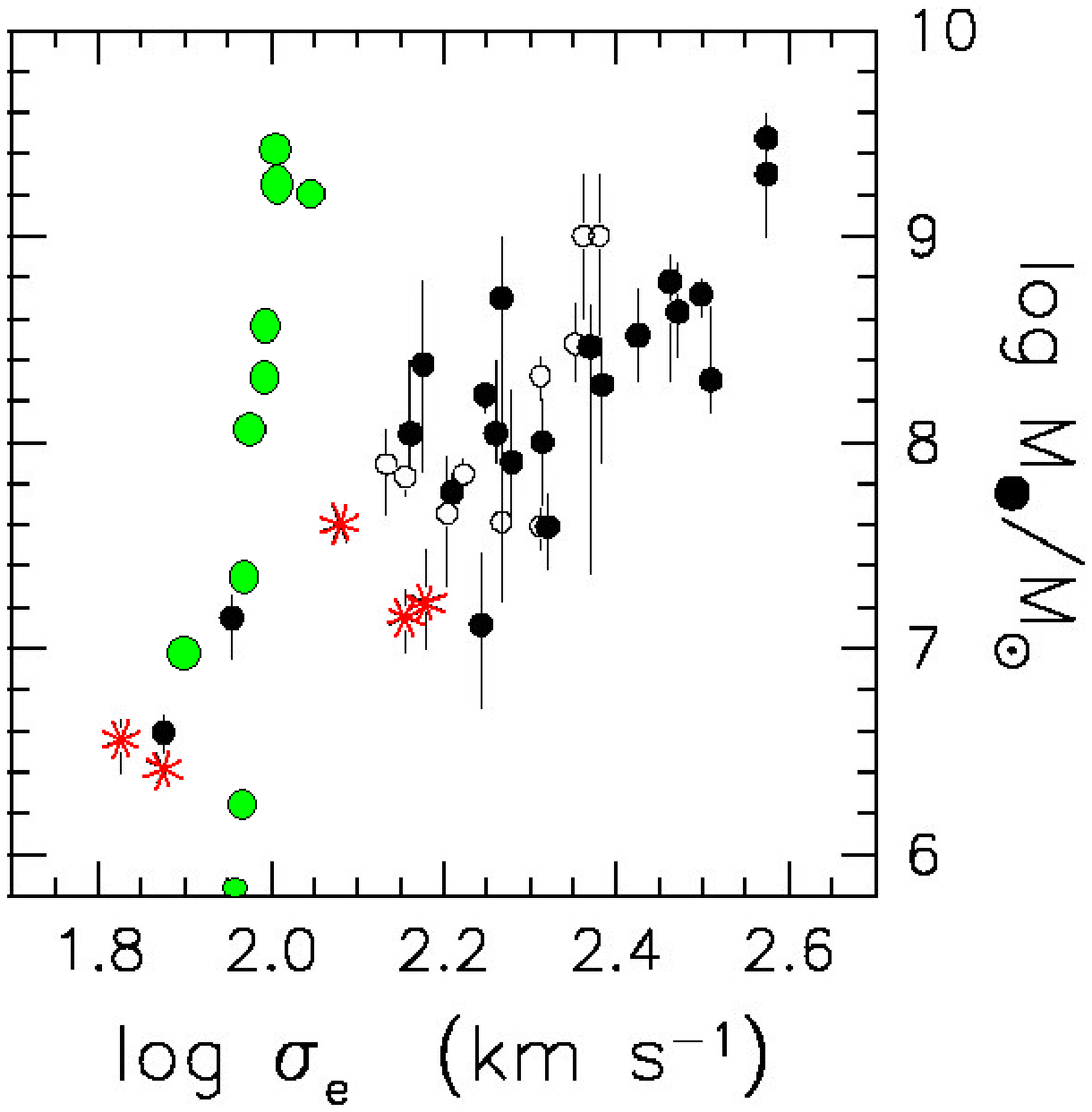}
%\includegraphics[width= 6cm]{8285f12b.ps}
%\end{center}
\caption{The BH masses versus the velocity dispersions in our
  simulations  (green points)  compared with the Kormendy observed data (cit)}
%(right panel).}
\label{bulge}
\end{figure}

Figure \ref{bulge} shows the values of the stellar velocity dispersion
inside the bulges at the end of the evolution as function of the BH masses.\\

We compared data from our simulations with observations by  \citet{Kor}. Data from our
simulations  are located in the pseudo-bulges region, with is consistent with
the origin of the bulge structure in our simulations: they are not classical
hot bulges, rather, they derive from secular evolution of the bar.  

We can't find the phenomenologically observed correlation of velocity
dispersions with the BH masses. It must be noticed that we
vary the BH mass and feedback parameter within the {\it same} DM halo,
so from this point of view our Fig. 8 simply suggest that varying the
feedback efficiency and the initial BH mass has no effect on the
stellar velocity dispersion, which remains a reasonable proxy for the
halo mass.

The values that better agree with observation are obtained from
simulations using a high BH feedback parameter.  This trend suggests an useful constraint for such parameter
in the simulations. Another possibility is that pseudo-bulges could
behave differently, in this regard, from hot bulges. This point
remains to be deepened both observationally and theoretically.

\section{Discussion and conclusions}

We have presented 11 isolated simulations and 2 cosmological simulations of the
evolution of a stellar+gaseous disk embedded in a DM halo,with the same
disk--to--halo mass ratios. The aim was  to evaluate the impact of a massive
black hole, with different accretion histories on the formation and evolution
of a stellar bar.

The old star component shows a long-lasting bar, 10 Gyrs old, in all our
simulations, regardless of the presence, the mass and the energy feedback
efficiency of a central BH.

We noticed a mild impact of the BHs having masses greater than $10^8 h^{-1}
M_{\odot}$  on the old star component ellipticity, impact
which appear stronger on the bar ellipticity of the new
stars. In a previous paper (\citet{Cu08}) we found that the star formation, by
reducing the central gaseous mass concentration, allows the bar to survive
until the end of the evolution in massive disks, at variance with the
results in \citet{Cu07} ( where the gas was not allowed to form stars): in
the latter case a gas fraction 0.2 was able to destroy the bar.

In this work we show that a BH placed and growing at the
center of the disk has a small influence on the ellipticity of the bar
developing in the preexistent star population.  On the other hand, the BH with its feedback has an action on the
formation of the CMC, which has a major role in quenching the bar. The bar formed by the newly formed stars is
always weaker than in the disk without BH, independently on the BH's evolution and
feedback. Also in this case, however, the presence of the BH does not
completely stop the bar formation.

  A direct comparison between the results of  our  $N$-body
  simulations and  the
  results of \citet{Hoz05} is not possible. The reason is that  their numerical
  model consists in  a razor-thin disk model, without bulge and halo. It is
evolved using a self-consistent field method (\citet{Hern92}), and not with
a $N$-body code.  Whereas the Hozumi's disk is stabilized by a black hole having
mass of $ 10^{8.5} M_{\odot}$ our disk forms a bar even if the central BH has
a higher mass.

 Adding gas and star formation to the same disk, we
notice that an accreting massive BH has an impact on the bar
ellipticity but it is not able to destroy it completely.

The effect of the BH is more evident  on the {\it inner} bar
ellipticity, which is reduced also when the effect of a star-forming
gaseous component is considered.
 
A comparison with the phenomenology confirms that the default feedback
parameter value 0.05 seems the more appropriate to reproduce the correlations
between the velocity dispersion and the black holes masses observed by
Kormendy.

{\bf Acknowledgments}  
Simulations were performed on the CINECA IBM CLX cluster, thanks to the
INAF-CINECA grants cnato43a/inato003 `` Simulations of disk galaxies in a
cosmological framework: the impact of the central Black Hole''. We wish to
thank V. Springel for kindly providing us with his code GADGET. We thank an
anonymous referee for his/her valuable suggestions, helpful to improve the paper.
\bibliographystyle{aa}
\bibliography{data.bib}
\end{document}